\documentclass[a4paper,10pt]{article}
\usepackage{graphics}
\usepackage{graphicx}
\usepackage{subfigure}
\usepackage{epsfig}
\usepackage{amssymb}

\title{Silicon nanostructures toxicity. An ab inito approach.}

\author{E.R. L. Loustau$^{ac}$, J. Klapp$^{ab}$}

\begin{document}

\maketitle

\begin{center}
$^{a}$Centro de Investigaci\'on y de Estudios Avanzados del Instituto Polit\'ecnico Nacional.\\
A. P. 14740, CP. 07360, D. F., M\'exico\\
$^{b}$Instituto Nacional de Investigaciones Nucleares\\. CP.52750, Toluca, M\'exico\\
$^{c}$Centro de Ciencias de la Complejidad. Universidad Nacional Aut\'onoma de M\'exico\\ A. P.70472, CP.04510, D.F., M\'exico.\\
\end{center}

\begin{abstract}
Silicon nanoparticles are widely used in the medical area and until now they have not manifested toxicological effects in humans beings. In order to understand the physical properties that determine their low-toxicity, we perform ab initio computational simulations of silicon nanoclusters, pure, p-doped and hollow structured. The topological and electronic properties obtained pointed out to the number of dangling bonds and electronic density as fundamental parameters to locate active sites and directly related to toxicity, besides the size and surface chemistry of the silicon nanoparticle.

\end{abstract}

\section{Introduction}

Silicon is the second more abundant element on the earth surface and is one of the most useful elements to mankind. Is best known for dominate the electronics industry, although is used too in the silicon ingot manufacture, solar cells, silicone and construction industries. The capacity to manipulate different structures of silicon like the crystalline, amorphous or porous it has allowed the fabrication of a wide variety of silicon devices. Nanotechnology has drawn attention to the properties of the silicon nanoparticles (NPSi) since they promise new and exiting applications. NPSi are defined as silicon structures with diameters less than 100nm and a surface to volume ratio bigger than ($\frac{60 m2}{cm3}$). Until now they have been proposed as vehicles for nutrients delivery in the human gut or as food storages because of its surprisingly stability and biodegradability \cite{Cahnam}. In cell biology and medicine their relevant photophysical properties suggest that they can be used as multimodal agents in bioimaging; especially as fluorescent labels that monitor changes in living cells, overcoming the photodegradation problem of the conventional fluorescent dyes \cite{Farrell}. Optimal physicochemical properties of nanoporous NPSi, as its excellent loading capacity with no chemical modification on the adsorbed drug, can be exploited in order to deliver drugs in humans efficiently \cite{Perrone} for example as a part of the cancer therapy \cite{Sazan}.\\

Despite the NPSi virtues, nanotoxicity of these particles should be characterized and understood before an intensive use on humans or, before released them in the environment. Examples of toxic effects of some nanoparticles include tissues inflammation, and altered cellular redox balance toward the oxidation causing the cellular death. The key to understanding the nanotoxicity is that their minute size, smaller than cells and cellular organelles, allows them to penetrate these basic biological structures, disrupting their normal function. Other important factors to the nanotoxicity are the nanoparticle morphology, chemical composition, surface chemistry, surface charge, porosity, aggregation state and agglomeration rate \cite{Buzea}. The toxicity of silicon nanoparticles is a controversial issue. The work of \cite{Polloth} mentioned that until now, there are not toxic effect on the human health of the synthetic amorphous silica nanostructures but, cytotoxic effects of 20nm silica nanoparticles in kidney cells have been reported by \cite{Passagne}. Porous silicon nanoparticles, oxygen passivated surface, have been showed to be toxic for the intestinal epithelial cells \cite{Bimbo}. In both cases the toxicity is associated to the stress oxidative with up-production of reactive oxygen species besides the silica and porous silicon nanoparticles capacity of penetrate the cells. A detailed review about the \emph{in vitro} and \emph{in vivo} adverse health effects of silica nanoparticles warns that most of the studies use poorly particles characterization methods in terms of their composition and physicochemical properties \cite{Napierska}. Therefore, the \emph{ab initio} simulations that can be carried out with the high computing power, allow better and more realistic models that could be helpful to address potentials and risks in silicon nanoparticles \cite{Barnard}.\\

Because physicochemical properties of nanoparticles determine their interaction with the cell and within the cell, and even subtle differences in such properties can modulate their toxicity and modes of action, this work is focus on chemical composition and atomic structure aspects of the NPSi. The surface of our NPSi models are oxygen free, in order to eliminate the oxygen reactive sites but they are passivated with hydrogen. One of our models contain a boron impurity because this kind of impurity remains after the electrochemical attack of p-type silicon wafers \cite{Polisski}, the most common fabrication method of silicon nanoparticles. In order to understand the influence of the structure on the NPSi toxicity, one of our models is hollow. Finally, since the NPSi applications take place at room or higher temperatures, we performed a molecular dynamics process on one of the clusters. On the method section we mentioned all the computational parameters, the visualization-softwares and the analysis tools used. Some topological and electronic properties of the clusters are reported on the results section. General observations and remarks are presented on conclusions.

\section{Method}
We generate 4 different clusters of silicon named NPSi, NPSiMD, NPSi type P and HollowNPSi. The optimization and molecular dynamics process were performed with the PWscf code implemented on the Quantum Expresso package, parallel version 5. 0. 1. \cite{QE}. The PWscf code is based on the density functional theory (DFT) (\cite{Hohenberg}, \cite{Sham}), uses plane waves to expand the electronic wave functions, pseudopotentials to simulate the electronic behavior and self consistent field steps to calculate the electronic density value. The Si.pz-vbc.UPF, H.pz-kjpaw.UPF and B.pz-n-kjpaw-psl.0.1.UPF potentials were chosen for silicon, hydrogen and boron atoms respectively. The wave function cut off radius was 50 Ry and 200 Ry for the electronic density as suggested by the potentials files.

\begin{figure}
\begin{center}
\subfigure[NPSi]{\label{NPSipure}\includegraphics[width=0.4\textwidth, height=4.0 cm]{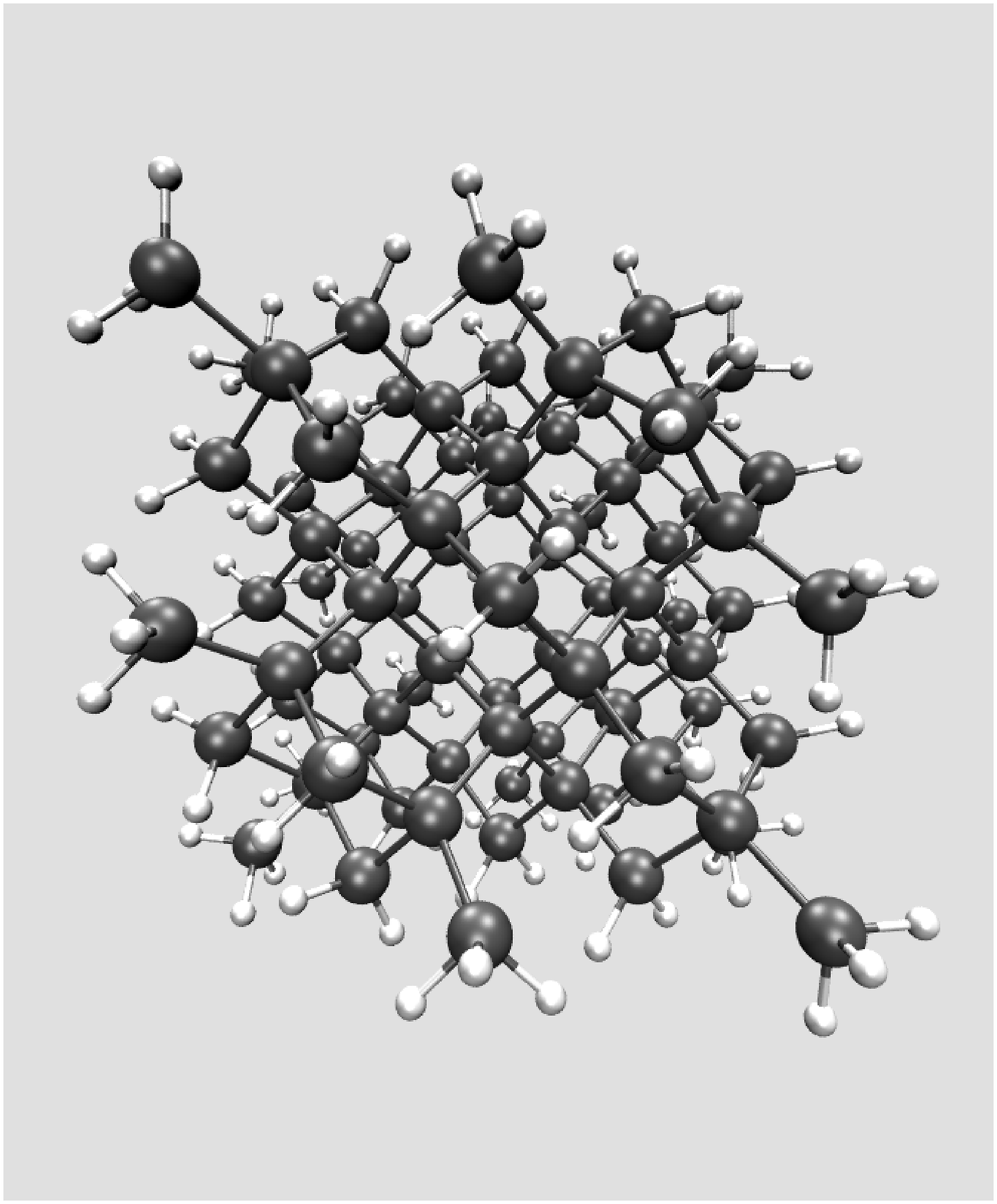}}
\subfigure[NPSi with MD]{\label{NPSiMD}\includegraphics[width=0.4\textwidth, height=4.0 cm]{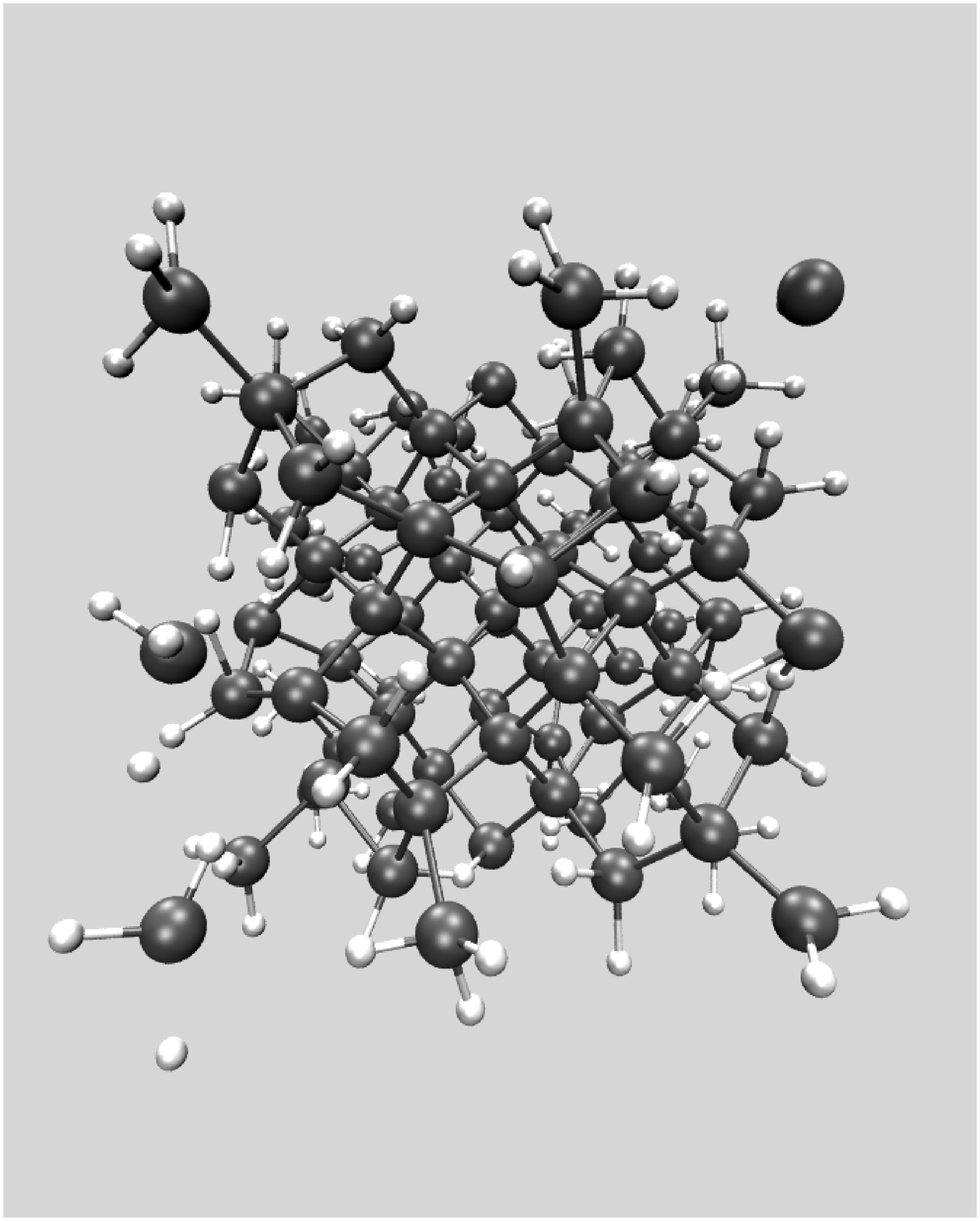}}
\subfigure[NPSi Type P]{\label{NPSiB}\includegraphics[width=0.4\textwidth, height=4.0 cm]{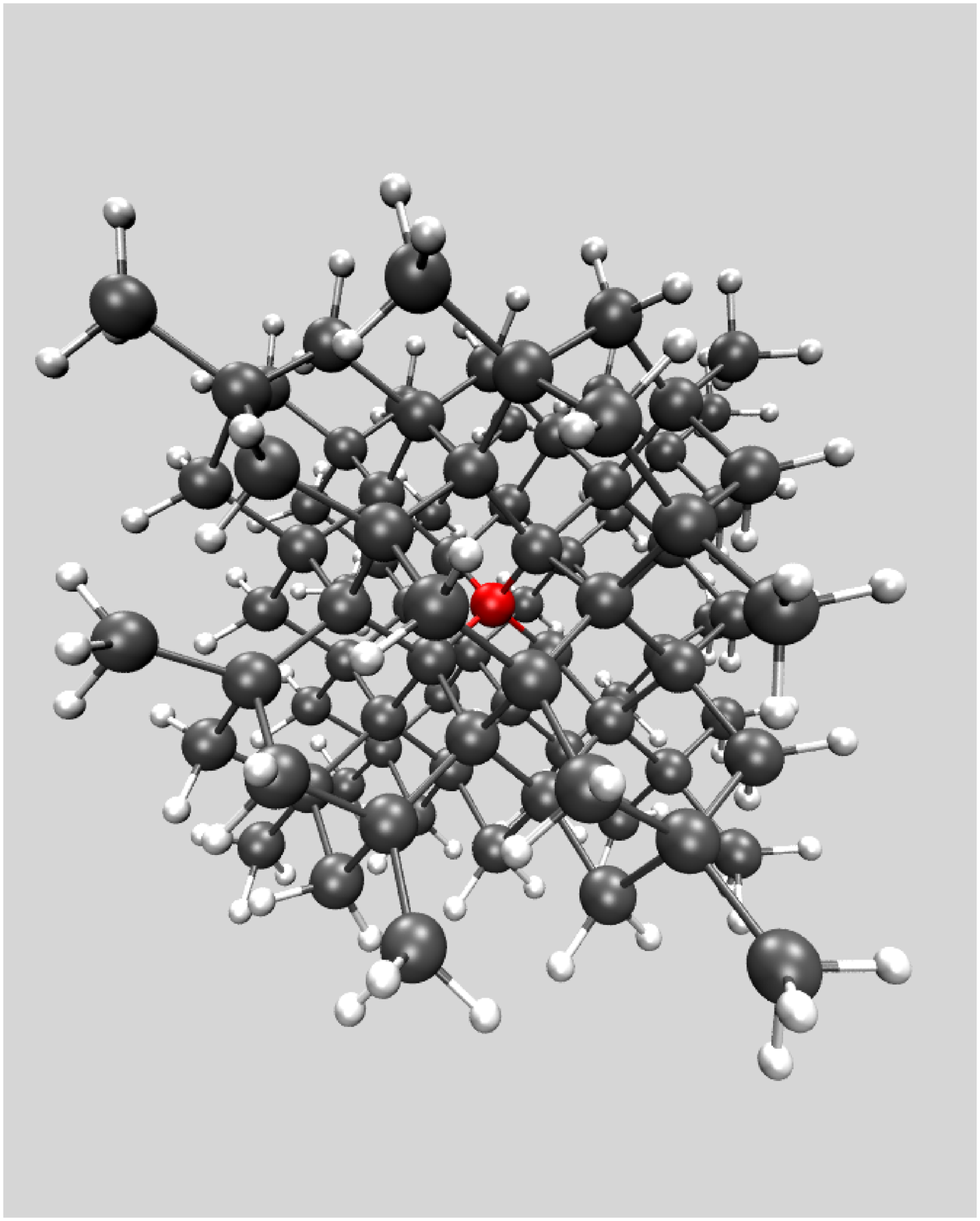}}
\subfigure[Hollow NPSi]{\label{HollowNPSi}\includegraphics[width=0.4\textwidth, height=4.0 cm]{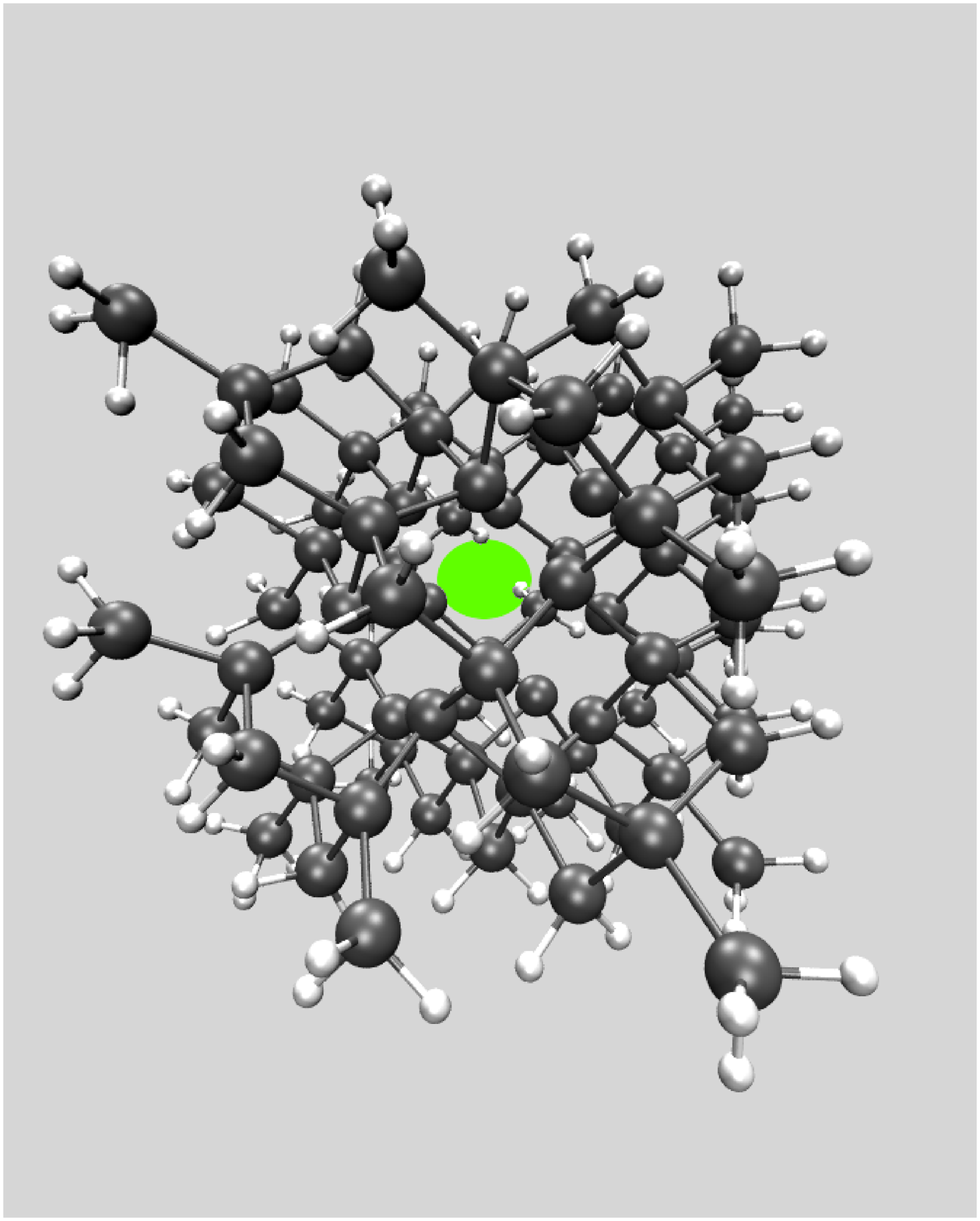}}
\caption{\label{Clusters}. (a) Optimized structure of NPSi, (b) Optimized structure of NPSi after a molecular dynamics process, (c) the boron (red atom) doped structure of NPSi and (d) a hollow (green sphere) NPSi structure.}
\end{center}
\end{figure}

\begin{enumerate}

 \item The diamond structure \textbf{Si} cell was replicated 2 times on each axis to obtain a crystalline supercell with 64 \textbf{Si} atoms, whose edges length were 10.86 {\AA} and with a density of 2.33 $gr/cm^{3}$. To generate a silicon nanoparticle instead of a periodic system, we had to consider the silicon surface atoms of the supercell. Once the distance bond was checked, the silicon dangling bonds were passivated with hydrogen. The seed structure was composed by 91 silicon and 108 hydrogen atoms.

 \item  The optimization of the seed structure produced the NPSi cluster \ref{NPSipure}.

 \item  Then, we applied a molecular dynamics process during 200 time steps, at 300 K, on the NPSi optimized structure to generate the NPSiMD cluster. Each time step was 20 fs long \ref{NPSiMD}.

 \item The central silicon atom of the optimized NPSi cluster was removed and replaced by a boron atom to generate the NPSi type P model. Since the valence of boron is 3, one of the 4 bonds visualized does not exist \ref{NPSiB}.

 \item A sphere of 3 {\AA} radius was considered from the center of the NPSi optimized structure, and the 5 silicon atoms within the sphere where removed. Then, the HollowNPSi model with 86 silicon and 108 hydrogen atoms was optimized \ref{HollowNPSi}.

 \end{enumerate}

The VMD and XCRYSDen softwares were used to visualize the atomic models and the ISSACS code to obtain their radial distribution functions, plane angle and dihedral distributions.

\section{Results}
\begin{figure}
\begin{center}
\subfigure[Radial distribution functions]{\label{RDF}\includegraphics[width=0.6\textwidth, height=6.0 cm]{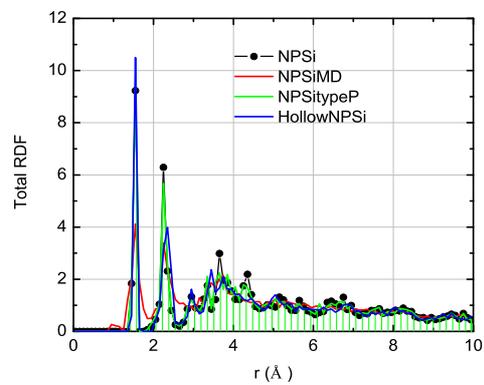}}
\subfigure[Plane Angles]{\label{Planos}\includegraphics[width=0.6\textwidth, height=6.0 cm]{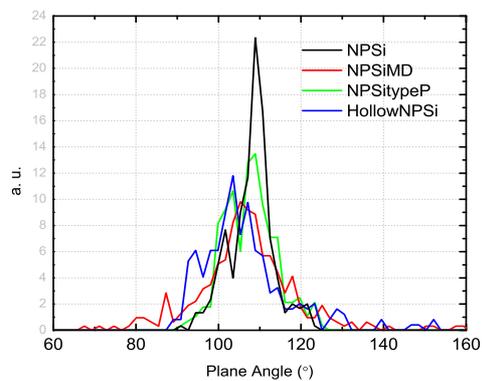}}
\subfigure[Dihedral Angles]{\label{Diedrales}\includegraphics[width=0.6\textwidth, height=6.0 cm]{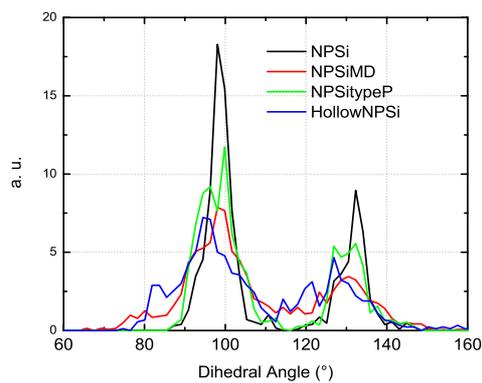}}
\caption{\label{Topology}. (a) Total radial distribution functions, (b) Plane angle distributions and (c) Dihedral angle distributions of the NPSi models.}
\end{center}
\end{figure}

In order to obtain information about the structures at small range order (less than 5 {\AA}) we calculated their plane angles distributions and coordination numbers. Medium range order (5-20 {\AA}) can be analyzed by their dihedral distributions, and the
radial distribution functions (RDFs) give a global view of the structures network behavior (figure \ref{Topology}) \cite{Elliot}. The RDFs of the NPSi and NPSi type P clusters are similar: 2 sharp picks and a third region corresponding to the second silicon neighbors. In contrast, the RDFs of the NPSiMD and HollowNPSi clusters present only two broad picks implying more disordered structures. For the NPSi and NPSi type P models, the most probable interatomic distance Si-H is on the range of $1.56-1.65$ {\AA}, the first silicon neighbors are bonded from 2.25 to 2.56 {\AA} and the second silicon neighbors can be found around 3.65 {\AA}. The HollowNPSi model increased only 8 \% the Si-H distance and maintain the same separation between the silicon atoms than the previous models. The bonding distance Si-H, and the first silicon neighbors separation are increased by 12 \% and 16 \% respectively on the NPSiMD model \ref{RDF}. The plane angle distributions are presented in figure \ref{Planos} and as can be noted, the NPSi and NPSitypeP distributions are sharper than the corresponding to the NPSiMD and HollowNPSi models, pointed out a more crystalline behavior. Most of the plane angles of the NPSi and NPSi type P models are around 110 degrees, near the silicon crystalline plane angle value of 109.5 degrees ($(\theta)_c$), although some atoms are bonded around 100 and 120 degrees. The broad distributions of the NPSiMD and HollowNPSi models are centered around $(\theta)_c$ but the height of their major picks reveals that less silicon atoms are bonded at an angle near $(\theta)_c$ than in the pure and doped models. According to \cite{Yoo}, the existence of endohedral (internal atoms bonded in a tetrahedral) silicon atoms and sphere like configuration of our silicon clusters are the consequence of a silicon bulk-like potential choice. The interatomic distance and plane angle distributions of our clusters describe correctly a more crystalline behavior for the pure and doped models than for the NPSiMD and HollowNPSi clusters. The form of the RDFs obtained by the NPSiMD and HollowNPSi models suggest the ones reported on \cite{Fernando} where an amorphous silicon network was studied. The coordination numbers of our models were 3.6 for the NPSi and NPSi doped models, and 3.9 and 3.7 for the NPSiMD and HollowNPSi models respectively. Then, we can conclude that the room temperature applied during the MD process promotes the surface reconstruction decreasing a the same time the number of dangling bonds. Our observations are consistent with the ones reported on \cite{Galashev}, where room and higher temperatures were applied on big silicon clusters generating amorphous and vitreous structures. Finally, the hollow model seems to become amorphous after the optimization process indicating that, the porosity of the silicon nanoparticles could determine its degree of crystallinity \cite{MyPorosity}. Dihedral angle describes the relative orientation of adjacent tetrahedral \cite{Zallen}. Since the diameter of our NPSi models is about 11 {\AA}, the dihedral distributions accounted the orientation of adjacent tetrahedral lying on the external layers or surfaces of the clusters but not in their centers. The distributions  (figure \ref{Diedrales}) showed a major pick near 90 degrees and a second pick around 130 degrees, a clear deviation from the crystalline value of 60 degrees. The picks of the  NPSi and NPSi type P are sharper than the picks of the NPSiMD and HollowNPSi distributions, pointing the distortion of the crystalline network to the amorphous one, due to the room temperature and porosity factors.

The total energy per atom ($e_tot$) was calculated and we found that, the HollowNPSi $e_tot=4.18$ eV was the smallest and the biggest one was the $e_tot=4.39$ eV of the MDNPSi model. The $e_tot$ values of the NPSi and NPSitypeP models were between the extremes mentioned. The $e_tot$ of the NPSitypeP was 0.27 eV higher than the $e_tot$ of the pure model NPSi, because of the boron impurity in accordance with \cite{Iori}. The Fermi levels of the MDNPSi, HollowNPSi, NPSitypeP and NPSi models were -2.23, -3.09, 3.69 and -3.74 eV respectively.

\begin{figure}
\begin{center}
\subfigure[Doped NPSi]{\label{ELFNPSiB}\includegraphics[width=0.8\textwidth, height=6.0 cm]{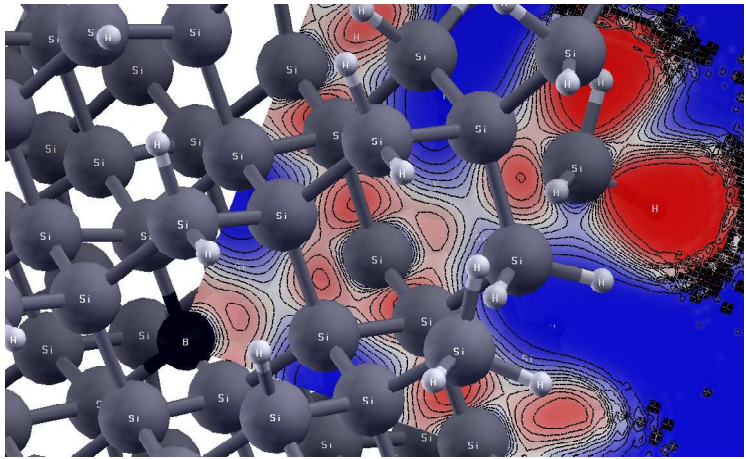}}
\subfigure[NPSi after the MD process]{\label{ELFNPSiMD}\includegraphics[width=0.8\textwidth, height=6.0 cm]{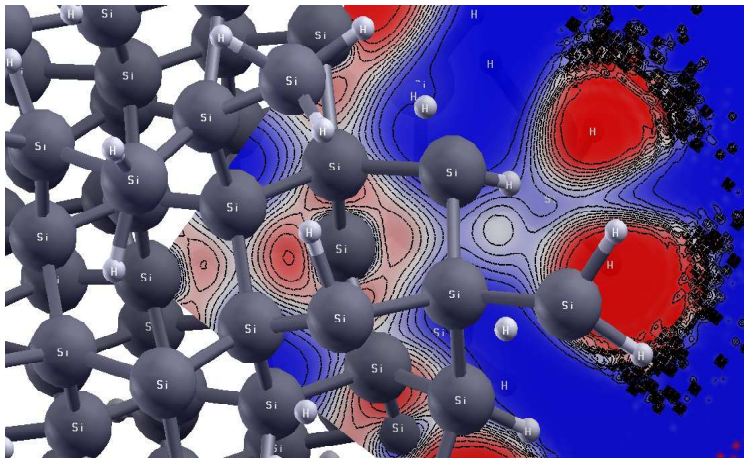}}
\subfigure[Hollow NPSi]{\label{ELFHollowNPSi}\includegraphics[width=0.8\textwidth, height=6.0 cm]{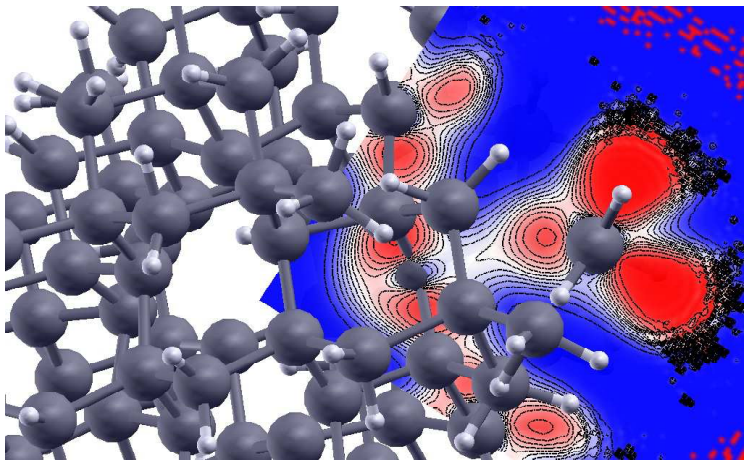}}
\caption{\label{ELFmodels}. Electron local functions (ELFs) of: (a) Doped NPSi model. The boron impurity is in black at the origin of the plane, (b) NPSi model after a molecular dynamics process. Silicon atom at the origin of the plane, (c) Hollow NPSi nanoparticle. Vacancy at the origin.}
\end{center}
\end{figure}

In order to investigate the electronic behavior of the models, their electron local functions (ELF) were obtained \ref{ELFmodels}. The ELF describes the probability of finding an electron in the neighborhood space of a reference electron, located at a given point and with the same spin \cite{Becke}. Our ELFs were colored on a blue-white-red code in which red means the highest probability of finding an electron and the blue color the lowest one. The reference point was located at the center of the models where a silicon, a boron atom and a vacancy lie in the NPSiMD, NPSi type P and HollowNPSi nanoparticles respectively. As we can see from figures \ref{ELFNPSiB}, \ref{ELFNPSiMD} and \ref{ELFHollowNPSi} the highest ELF are found around the hydrogen atoms; the ELF around the boron impurity is higher than the corresponding to a silicon atom because the boron orbital are much more localized than those of silicon in accordance with \cite{Chang}. Finally, there is not probability of finding and electron in the inner pore surface created by the vacancy.

\section{Conclusions}
From topological results we can conclude that a crystalline nanoparticle can be converted by the room temperature to an amorphous one. A hollow on a nanoparticle may cause a degree of amorphicity too. In the other hand, a boron doped nanosilicon particle remains essentially crystalline because the impurity modify the atomic network only locally. Because the number of dangling bonds decrease on the amorphous structures the reactive sites decrease as a result and then, the toxicity of these particles should be lower than the corresponding to the crystalline silicon nanoparticles. Electronic results indicate that the NPSi reactive sites are located on their cluster surfaces instead of, near substitutional boron impurities or inner pore surfaces. Because for the biological systems nanoparticles size is the principal toxic factor, bigger clusters NPSi simulations are under way in order to model more realistic topological and electronic properties.

\section{ACKNOWLEDGEMENTS}

Work supported by ABACUS, CONACyT grant EDOMEX-2011-C01-165873. Time computing from the New Supercomputing Cluster (NES) of the National Autonomous University of Mexico (UNAM). Dr. Jes\'us Antonio del R\'io for helpful discussion and computing facilities. Dr. Jos\'e Angel Reyes Retana for QE technical support and time computing.


\begin{thebibliography}{30}
\bibitem{Cahnam}
L.T. Canham. Nanoscale semiconducting silicon as a nutritional food additive. Nanotechnology18(2007)185704.

\bibitem{Farrell}
Norah O Farrel, Andrew Houlton, Banjamin R Horrocks. Silicon nanoparticles: applications in cell biology and medicine. International Journal of Nanomedicine 4(2006:1)451-472.

\bibitem{Perrone}
M. Perrone Donnorso, E. Miele, F. De Angelis, R. La Rocca, T. Limongi, F. Cella Zanacchi, S. Marras, R. Brescia, E. Di Fabrizio. Nanoporous silicon nanoparticles for drug delivery applications. Microelectronic Engineering 98(2012)626-629.

\bibitem{Sazan}
Sazan M. Haidary, Emma P. Córcoles, and Nihad K. Ali. Nanoporous Silicon as Drug Delivery Systems for Cancer Therapies. Journal of Nanomaterials, vol. 2012, Article ID 830503, 15 pages, 2012. doi:10.1155/2012/830503

\bibitem{Buzea}
Cristina Buzea, Ivan I. Pacheco Blandino and Kevin Robbie. Nanomaterials and nanoparticles: Sources and toxicity. Biointerphases, vol. 2, issue 4 (2007), pages MR17-MR172.

\bibitem{Polloth}
Claudia Fruijtier-Polloth. The toxicological mode of action and the safety of synthetic amorphous silica-A nanostructured material. Toxicology 294 (2012)61-79.

\bibitem{Passagne}
Isabelle Passagne, Marie Morille, Marine Rousset, Igor Pujalt\'e, Beatrice L'Azou. Implication of oxidative stress in size-dependent toxicity of silica nanoparticles in kidney cells. Toxicology 299(2012)112-124.

\bibitem{Bimbo}
Luis M. Bimbo, Ermei M$\ddot{a}kil\ddot{a}$, Timo Laaksonen, Vesa-Pekka Lehto, Jarno Salonen, Jouni Hirvonen, H\'elder A. Santos. drug permeation across intestinal epithelial cells using porous silicon nanoparticles. Biomaterials 32(2011)2625-2633.

\bibitem{Napierska}
Dorota Napierska, Leen CJ Thomassen, Dominique Lison, Johan A. Martens, Peter H. Hoet. The nanosilica hazard: another variable entity. Particle and fibre toxicology 7(2010)1-39.

\bibitem{Barnard}
Amanda S. Barnard. How can \emph{ab initio} simulations address risks in nanotech?. Nature Nanotechnology 4(2009)332-335.

\bibitem{Polisski}
G. Polisski, D. Kovalev, G. Dollinger, T. Sulima, F. Koch. Boron in mesoporous Si-Where have all the carriers gone? Physica B273-274(1999)951-954.



\bibitem{QE}
 www.quantum-expresso.org

\bibitem{Hohenberg}
         P. Hohenberg and W. Kohn, Phys. Rev. B. 136(1964)864-871.

\bibitem{Sham}
         W. Kohn and L. J. Sham, Phys. Rev. A. 140(1965)1133-1138.



\bibitem{Elliot}
S. R. Elliot. Physics of amorphous materials. Edit. Longman Group, Hong Kong, 1990, 73-75 and 132-139 p.p.

\bibitem{Yoo}
S. Yoo, Xiao Cheng Zeng. Global geometry optimization of silicon clusters described by three empirical potentials. J. Chem. Phys. Vol. 119, 3(2003)1442-1450.

\bibitem{Fernando}
Fernando Alvarez, Ariel A. Valladares. Ab initio generation of amorphous semiconducting structures. The case of a-Si. Journal of Non-Crystalline Solids 229-302(2002)259-264.

\bibitem{Galashev}
A. E. Galashev, I. A. Izmodenov, A. N. Novruzov and O. A. Novruzova. Computer Study of Physical Properties of Silicon Nanostructures. Semiconductors, Vol. 41, 2(2007)190-196.

\bibitem{MyPorosity}
Emilye R. L. Loustau and Ariel A. Valladares. Crystalline and amophous nanostructures in porous silicon. Journal of Non-Crystalline Solids, Vol.354,19-25(2008)2200-2203.

\bibitem{Zallen}
Richard Zallen. The physics of amorphous solids. Edit. John Wiley and Sons, NY, 1998, 69-70 p.p.

\bibitem{Iori}
F. Iori, S. Ossicini. Effects of simultaneous doping with boron and phosphorous on the structural electronic and optical properties of silicon nanostructures. Physica E 41 (2009)939-946.

\bibitem{Becke}
A. D. Becke and K. E. Edgecombe. A simple measure of electron localization in atomic and molecular systems. Journal of Chemical Physics 92, 9(1990)5397-5403.

\bibitem{Chang}
Jianlin Chang and M. J. Stott. Si(001)$/$B surface reconstruction. Physical Review B53, 20(1996)13700-13704.

\end{thebibliography}
\end{document}